\documentclass{aa}
\usepackage[dvips]{graphics}

\newcommand{\Ha}{H$\alpha$}

\newcommand {\Na} {NaD$_2$}

\begin{document}
\thesaurus{09{09.16.2; 09.03.1; 09.13.1; 09.15.1}} 
\title{ Network and internetwork: a compared multiwavelength analysis}
\titlerunning{Network and internetwork}

\author{G. Cauzzi\inst{1}
\and A. Falchi\inst{1}
\and R. Falciani\inst{2}}

\offprints{A. Falchi, e-mail: falchi@arcetri.astro.it}

\institute{Osservatorio Astrofisico di Arcetri, %
       I-50125 Firenze, Italy
\and Dip. di Astronomia e Scienza dello Spazio, %
       I-50125 Firenze, Italy}

\date{Received 23 december 1999 /Accepted 16 march 2000}
\maketitle

\begin{abstract}
We analyze the temporal behavior of Network Bright Points (NBPs)
using a set of data acquired during
coordinated observations between ground-based
observatories (mainly at the NSO/Sacramento Peak) and the
Michelson Doppler Interferometer onboard 
SOHO. 

We find that, at any time during the observational sequence,
all the NBPs visible in the NaD$_2$ images
are co-spatial within 1$^{\prime\prime}$ with locations of enhanced magnetic 
field. The ``excess'' of NaD$_2$ intensity in NBPs, i.e. the emission 
over the average value of quiet regions, is directly related to the
magnetic flux density. This property implies that, in analogy with the Ca
II K line, the NaD$_2$ line center
emission can be used as a proxy for magnetic structures.

 We also compare the oscillation properties of NBPs and internetwork areas. At
photospheric levels no differences between the two structures
are found in power spectra, but analysis of phase and coherence spectra
suggests the presence of downward propagating waves in the internetwork.
At chromospheric levels some differences are evident in the power
spectrum between NBPs and internetwork. At
levels contributing to the \Na~emission the NBPs show a strongly
reduced amplitude of oscillations at the $p$ - mode frequencies.
At levels contributing to the \Ha \ core emission, 
the amplitude of network oscillations is higher
than the internetwork ones. The power spectrum of NBPs at this wavelength 
shows an important peak
at 2.2 mHz (7 minutes), not present in the internetwork areas. Its
coherence spectrum with \Ha \ wings shows very low coherence at this
frequency, implying that the oscillations at these chromospheric levels
are not directly coupled with those present in lower layers. 


\end{abstract}
\section{Introduction}
\label{s_int}

The chromospheric bright network has long been observed in narrowband
spectroheliograms taken in the H and K cores of the Ca II resonance
lines. The Ca II network typically shows H and K profiles with high double
peaks and enhanced line wings that persist for extended periods of 
time (longer than 10
minutes, see, e.g., Rutten \& Uitenbroek 1991 ). The chromospheric network 
emission pattern is cospatial with small-scale magnetic field
concentrations, and defines the supergranular network boundaries. 
It is this atmospheric component that
produces the correlation between H and K excess line-core flux and
magnetic activity of cool stars (\cite{schri89}).

The dynamics of the network
elements, compared with the internetwork or quiet chromosphere, has been
extensively studied (especially from the observational point of view)
since these small-scale structures can be important in
channeling the energy from photospheric layers to the transition region
and corona (\cite{kne86}, 1993, \cite{deu90}, \cite{kul92}, %
Al et al. 1998).

An assessment of the spectral characteristic properties of Network 
Bright Points (NBPs) at different layers in the atmosphere has been
provided by Lites et al. (1993) using
spectral observations in the range of the Ca II H line.
In their work, these authors analyzed spectrographic observations of a
single network bright patch and of several internetwork
points. The wavelength shifts of photospheric and chromospheric
lines allowed them to perform a compared analysis between the dynamics of 
the two atmospheric components. One of the relevant characteristics they
describe is that
at chromospheric levels (Ca II H$_3$) the NBPs show long period 
oscillations ($\nu <$ 3 mHz) not correlated
with oscillations in the lower atmosphere, while they lack power at higher
temporal frequencies. The internetwork regions
display instead enhanced power at higher frequencies, well correlated with
photospheric oscillations.
The presence of these low frequency oscillations in the network
has been confirmed by Lites (1994) also for the chromospheric He I 10830
line, in contrast to Bocchialini et al. (1994) which observe, for the same
line, oscillations only in the 5 minutes range. 

An enhanced
power in the low frequency range for network points with respect to the
internetwork has also been observed by
Kneer \& von Uexk\"{u}ll (1986) in the center of the chromospheric \Ha~
line. These authors however interpret this feature as not due to
oscillations, but of mainly stochastic origin, and attribute it to erratic 
motions of the corresponding photospheric footpoints.

The problem is still open, and further observations to better address 
this issue are required (\cite{lit94}). In
particular one would need observations: 1- on a larger number of NBPs, to
improve on the statistics; 2- at different heights in the atmosphere, since
the analysis of the coherence between fluctuations at different levels 
can help exploring the nature of oscillations.
To this end, a reliable method for the identification of the same physical
structure at different atmospheric levels is mandatory, since the
inclination of the magnetic field could displace the chromospheric network 
points with respect to the corresponding photospheric ones. 

In this paper, we address some of these issues, and present observational
results on the NBPs and internetwork characteristics as derived from a
multiwavelength analysis. The observations were obtained in
August 1996, during  a coordinated observing program between 
ground-based observatories and the
Solar and Heliospheric Observatory (SOHO).
For the ground-based observations we used the cluster of instruments at the
NSO/Sa\-cra\-men\-to Peak R.B. Dunn Solar Telescope (NSO/SP-DST), 
that could provide a complete coverage at lower atmospheric levels.
The dataset used is described in Sect. \ref{s_obs}.
General properties of a sample of NBPs, followed from the
photosphere up to the chromosphere and including their relationship with the
magnetic structures, are given in Sect. \ref{s_char}. The temporal
development of the NBPs is described in Sect. \ref{s_lcurves}. Sect.
\ref{s_power} and \ref{s_phase} provide an
analysis of the power, phase difference and coherence spectra for the
fluctuations observed separately within the NBPs and the surrounding
internetwork. Finally, discussion and conclusions 
are given in Sect. \ref{s_concl}.

\begin{table*}[ht]
\caption{Summary of the observations.  }\label{tbl-1}
\begin{center}
\scriptsize
\begin{tabular*}{18.2cm}{llllll}
Instrument & FOV & Spat. resol. & Observing $\lambda$ (\AA) & FWHM (\AA)
& $\Delta$t (s) \\
 & & & & & \\
UBF & 2$^\prime \times 2^\prime$ & $0.5^{\prime\prime} \times
0.5^{\prime\prime}$ & 5889.9 (NaD$_2$) & 0.2 & 12 \\
 & & & 6562.8 (H$\alpha$) & 0.25 & \\
 & & & 6561.3 (H$\alpha -1.5$ \AA) & 0.25 & \\
Zeiss & 2$^\prime \times 2^\prime$ & $0.5^{\prime\prime} \times
0.5^{\prime\prime}$ & 6564.3 (H$\alpha +1.5$ \AA) & 0.25 & 3 \\
White Light & 2$^\prime \times 2^\prime$ & $0.5^{\prime\prime} \times
0.5^{\prime\prime}$ & 5500 & 100 & 3 \\
HSG & $0.75^{\prime\prime}\times 2^\prime$ & $0.75^{\prime\prime}\times
0.36^{\prime\prime}$ & 3904$-$3941 (CaII K) & 0.035 &   \\
MDI &10$^\prime \times 6^\prime$ &$0.6^{\prime\prime}\times
0.6^{\prime\prime}$& 6768 (Ni I) & 0.1 &60 \\
\end{tabular*}
\end{center}
\end{table*}

\section{Observations and data reduction}
\label{s_obs}

A description of the general data acquisition has been  
presented in Cauzzi et al. (1997, 1999).
Table 1 gives a summary of the observing setup, and we recall here only
some short information on the data used in this paper.
Monochromatic intensity images were obtained with the tunable 
Universal Birefringent Filter (UBF) and the Zeiss filter, at high spatial 
and
temporal resolution. Several spectra were obtained around the 
chromospheric 
CaII K line with the Horizontal Spectrograph (HSG). The spectra have 
been acquired setting the spectrograph slit on different bright points at 
different times; the field of view in the HSG row of Table 1 hence
refers to a single slit exposure. 
Onboard SOHO, the Michelson Doppler Imager (MDI, 
\cite{scher95}) 
acquired data in high resolution mode, i.e. with an image scale of
0.605\arcsec/pixel. Maps of pseudo-continuum intensity,
line-of-sight velocity, and longitudinal magnetic flux were obtained in the NiI
6768 \AA~line at a 
rate of one per minute for several hours. The line-of-sight velocity images 
were available in a
binned 2x2 format, i.e. with an effective spatial scale of 1.2\arcsec/pixel.

\hyphenation{smo-oth-ing}
We observed the small Active Region (AR) NOAA 7984  
over 5 consecutive days (Aug. 15 - 19, 1996). 
Its activity was very low and although some stronger magnetic
structures (a small spot and some pores) were present in the field of
view (FOV), no major eruptions of magnetic flux were recorded. 
The situation was then appropriate to study
and characterize the properties of NBPs visible within and around the
AR.
We remark that our set of data allows us to directly compare the NBPs 
as visible at different wavelengths with the corresponding magnetic 
field structure, and to follow them 
 from the deep photosphere to the higher chromosphere.
In this paper we analyze the data obtained on August 15th, 1996, since for
that day we
had the best uniformity in time coverage for NSO and MDI data. 
Fig. 1 shows the FOV at several wavelengths.
The period of best seeing for ground-based observations ran from 
15:15 to 16:05 UT. This interval is adequate 
for the study of network points, since it 
allows  an analysis of their (possible) periodical properties,
while they still
maintain their identity (\cite{lit93}; \cite{kne95}). 
MDI data were available for many hours around this interval; we consider in this
work the period 14:00 - 17:00 UT. 

\begin{figure*} 
\resizebox{\hsize}{!}{\includegraphics{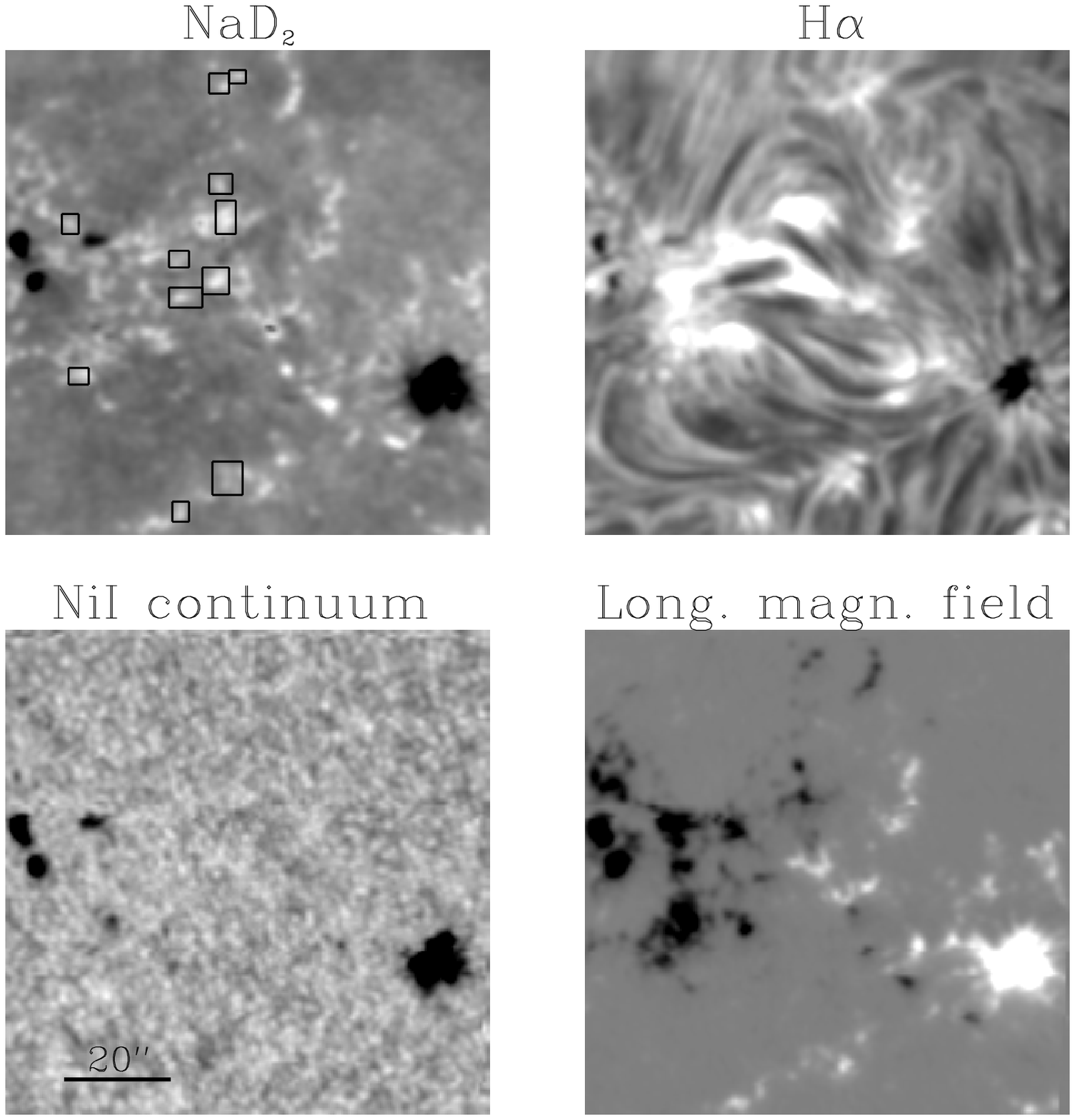}} 
\caption{ \label{fig1}AR NOAA 7984 observed on Aug. 15, 1996, N29E16.
a) NaD$_2$ image averaged  
over the time interval 15:15 - 16:05 UT. The black squares indicate the
position of the NBPs analyzed in this work. 
b), c) and d) Same FOV as a), observed in H$\alpha$, 
NiI pseudo-continuum and longitudinal magnetic flux.} 
\end{figure*}

The data were re-scaled to the 0.605\arcsec/pixel of the MDI maps, and the
co-alignment of the whole dataset was obtained comparing the position of 
the prominent solar features, 
i.e. the little spot and pores (see Fig. \ref{fig1}).
At each given time the alignment among  the images acquired with different
instruments was better than about  $1^{\prime\prime}$, the mean spatial
resolution limit of the ground-based frames. 

\section{Characteristics of NBPs }
\label{s_char}
In order to identify suitable network points, we first selected 
on the spectra bright features showing strong K$_2$
peaks and enhanced wings emission in CaII K. 
Since the NBPs lifetime is
typically longer than 10 minutes (Rutten \& Uitenbroek 1991),
we further required that the points be visible for the entire observing
period in the NaD$_2$ images as bright structures with intensity above the 
average. 
A total of 11 NBPs with the required characteristics were selected. They are
distributed over the FOV both near the center of the AR and away from it,
as can be seen in Fig. \ref{fig1}-a. 
Several more network points (or structures) are visible on the longitudinal 
magnetic flux maps, but in this work we  
limit our analysis to only these 11 points for which 
corresponding CaII K spectra are available.

\subsection{Intensity and velocity} \label{s_int_vel}

Within the areas enclosing the NBPs defined above, we tried to identify
the bright points at each wavelength by choosing an intensity (or other
signature) threshold
that could clearly separate them from their surroundings.

On white light and NiI pseudo-continuum images, an intensity
threshold cannot clearly discriminate between NBPs and other areas.
The contrast averaged 
over the spatial locations corresponding to the NBPs is of the
order of  1\%, i.e. smaller than the rms noise 
calculated in quiet areas (about 3.5\% and 2\% for the white light and
pseudo-continuum images, respectively). This is consistent 
with the observations of Topka
et al. (1997), that find low continuum intensity contrast in network
points with 
magnetic flux density smaller than $\approx$ 300 G 
(as typical of our points, see Sect.  \ref{s_mag}).

The line of sight velocities averaged over the NBP areas are small,
about 80 m/s downward with respect to the average values over the
quiet areas of the FOV. 
Since the standard deviation of the measurements is
about 200 m/s, for both NPBs and quiet areas,
it is not possible to define a velocity threshold that allows to
discriminate the NBPs within the FOV.
The small average red-shift is however in agreement with recent observations
by Solanki (1993) and Mart\'{\i}nez-Pillet et al. (1994).

The NBPs are instead well visible in the images acquired in the 
H$\alpha$ far wings and NaD$_2$ center, 
as sharp and isolated bright structures of comparable size 
(3\arcsec$-$4\arcsec wide). 
The bright points, as seen at these wavelengths, spatially coincide 
within the overlapping error of 1\arcsec. 
In the H$\alpha$ center images the morphology is less clear. The presence of
contiguous features, at times brighter than the selected NBPs, 
made more uncertain their identification. In total, however, 10 of the 
11 considered NBPs  were unambiguosly identified in the H$\alpha$ center
images. %

The characteristics of the NBPs for different signatures, including their
typical contrast with respect to quiet areas,  are summarized
in Table \ref{tab2}. It must be remarked that the formation height of
these signatures has been computed in a mean quiet atmosphere,
i.e. that it represents only a {\it generic} indication for magnetic
structures such as the NBPs. 
In particular, due to the Wilson depression, the radiation coming from
photospheric
magnetic structures is believed to be formed in deeper geometric layers with
respect to non-magnetic ones. Since the spatial resolution for our 
observations is not sufficient to resolve the (supposedly)
elementary magnetic fluxtubes (with dimensions smaller than 0.3\arcsec, 
as seen for example in G-band images), the signals we
analyze are a non-linear combination of magnetic and non-magnetic ones. %
\begin{table}
\caption{\label{tab2} Characteristics of the NBPs at different atmospheric heights.
The first column gives the formation height for each wavelength,
 computed in the quiet
Sun VALC model (\cite{ver81}), taking into account the different
widths of the filters. The second column gives the typical 
diameter of the structures when they are clearly visible. The third column
gives the intensity contrast values, averaged over time. A range of values is
reported when variations are large over the sample. %
}
\scriptsize
\begin{tabular}{llll}
Observed features &  height(km) & size(\arcsec) &$\Delta {\rm I} / {\rm I} $\\
 & & &   \\
White Light & 0 & - & .01 \\
H$\alpha$ wings & 100. & 3 & .04 \\
Pseudo-continuum (Ni I ) & 100-200. & - & .01 \\
NaD$_2$ & 600. & 3 - 4 & .1 \\
 H$\alpha$ center & 1500. & 4 & .1 - .2 \\
   & & &  \\

\end{tabular} 
\end{table} %

\subsection{Magnetic structures} 
\label{s_mag}

The selected NBPs are clearly recognizable on the MDI magnetic 
maps as sharp and isolated structures, 3\arcsec$-$4\arcsec~ wide. 
Each one
corresponds to a patch of definite polarity, with magnetic flux densities 
ranging from a minimum of 30 G, to a maximum of about 250 G (for comparison, 
the
spot in the FOV has a maximum flux density of 1100 G). The error on a single 
pixel in each image
is given at about 15 G (\cite{schri97}). 
Since the network fields are mostly vertical (see, e.g., \cite{lit99}), the flux
measure is only slightly affected by the position of the FOV on the solar disk
($\cos\theta \sim 0.9$).

The magnetic evolution of the
NBPs is quite varied. Some of the points maintain stable
positions and flux values, while others experience a steady 
increase during the observing period. In one single case we see a weak
magnetic structure appearing during the observing sequence, 
simultaneously with the appearance of a network bright point.

The spatial correspondence between the NaD$_2$ network points and the  magnetic
structures is very good, and will be analyzed in more detail in the next
section. The same correspondence is noticeable also 
for the network points as seen in the H$\alpha$ wings,
although this is less evident than for NaD$_2$ 
due to their lower contrast.

\subsection{Magnetic structure and NaD$_2$ emission} 
\label{s_magd2}

We checked the spatial correspondence between the network points as 
visible in NaD$_2$ and in magnetic maps.  
To this end 
we first removed from both signatures, by means of appropriate smoothing
techniques, short term variations due to noise
and oscillations (especially in the 5-minutes range).
We then subtracted a
threshold, chosen as the average quiet area value for the NaD$_2$ 
images, and as the nominal data noise (15 G) for the magnetic ones.
Finally, within the 11 selected areas, a NBP was identified in both signatures
as the locus of the pixels exceeding 50\% of the local maximum. This 
allowed its clear separation from the surroundings. %

\begin{figure*}  %
\resizebox{12cm}{!}{\includegraphics{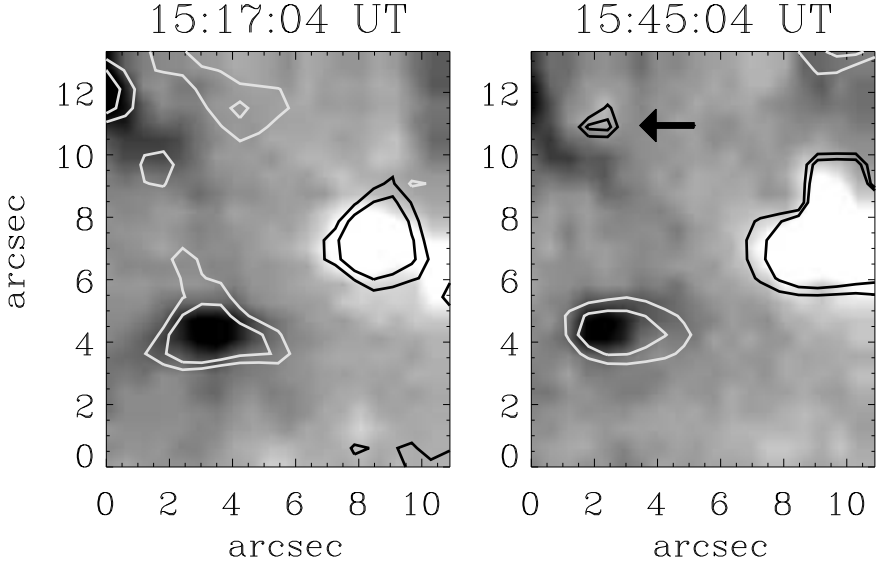}}
\hfill
\caption{
Longitudinal magnetic flux maps (scaled between $+$100 and $-$200 G) for 
two different NBPs at two different
times. The contours represent the NaD$_2$ threshold defining a network bright point.
One can see the change in shape and position of the magnetic 
structures during this interval, and the
corresponding changes in NaD$_2$ emission. The arrow indicates a new
NBP appeared simultaneously with a weak magnetic structure.} %
\label{figna_cm}
\end{figure*}

We find that, at {\it any} given time, the NaD$_2$ network points are 
coincident 
in position, size and shape with the corresponding magnetic patches, 
within 1\arcsec~(the overlapping error). %
Any change in the 
characteristics of the NaD$_2$ NBPs reflect almost perfectly those of 
the magnetic features,  within 
the temporal resolution of the MDI data. This is well
exemplified in Fig. 
\ref{figna_cm}, where we show NaD$_2$ contours overlaying magnetic flux maps
of several NBPs at two different times. 
To our knowledge, this is the first time
that such a correspondence is reported at high spatial {\it and} temporal
resolution.
A good agreement between the chromospheric
network emission pattern and the locations of 
enhanced magnetic flux had been noted in earlier works
(\cite{sku75}; \cite{schri89}; Nindos \& Zirin 1998 ) but mostly 
for the CaII K emission, with lower spatial and temporal resolution. 
No temporal resolution was available in the observations of Beckers (1976) 
or in those of Daras-Papamargaritis \& Koutchmy (1983), that established
a correlation between
magnetic structures and facular structures in the wings of the Mgb lines.

In analogy with the CaII H and K case, we could establish a {\it
quantitative} relationship between the NaD$_2$ excess and the magnetic flux
density for the network points. This property implies that also 
the emission in the center of the NaD$_2$ line 
can be used as a proxy for the magnetic field structures. For sake of
simplicity, the details of the determination of this relationship are given
in  \ref{s_app}.

\section{Temporal development: light curves}
\label{s_lcurves}

To study the temporal development of the NBPs, we computed the light curves for
the bright points at each wavelength or signature. The curves for each NBP were
obtained by selecting an area that contained the bright point throughout the 
whole observing period (even if it moved spatially), and then
averaging, for each time, over all the pixels whose intensity  exceeded
the threshold value described in Sect. \ref{s_int_vel}. A threshold equal to the
nominal data noise was used for the magnetic curves. 
As said in Sect. \ref{s_int_vel}, the NBPs are not directly visible in
some photospheric signatures, such as
white light, NiI pseudo-continuum and velocity images, hence we couldn't use an
intensity (velocity) threshold to obtain the corresponding light curves. To
guarantee the comparability with the other light curves, in
these cases we computed, at each time,
the average value over the spatially corresponding {\it magnetic} areas.

\begin{figure}  %
\resizebox{8.8cm}{!}{\includegraphics{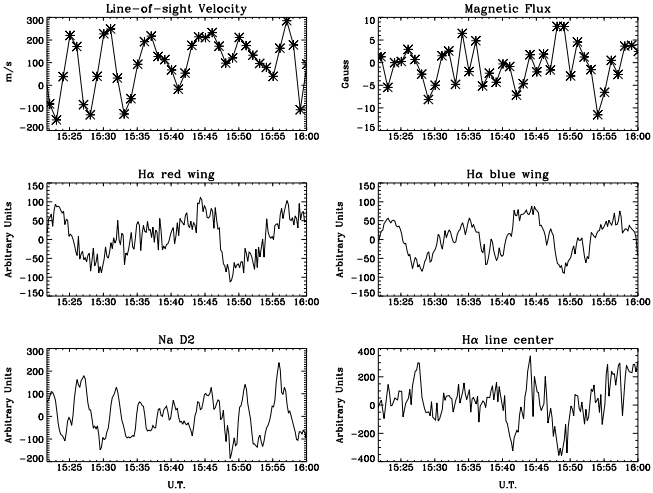}}
\caption{
Temporal evolution for a single NBP 
of velocity, magnetic flux density, intensity in H$\alpha$ wings, 
NaD$_2$, H$\alpha$. Long-period trends have been subtracted using a smoothing
window of 600 s (see next section). Intensity variations are  after
subtracting a value of 10$^4$ arbitrary
units} \label{fig_cur}
\end{figure}

We also computed the light curves of 11 areas 
randomly selected in the quiet regions of our FOV, and of size comparable to
that of the NBPs (about 3\arcsec$\times$3\arcsec). No threshold was applied 
for their computation. 
These quiet areas should represent the so-called internetwork
regions, which appear field-free at MDI sensitivity.  
Using the same number of internetwork areas as of network
bright points, and performing the analysis in the same way, will 
give us confidence in the comparison in a statistical sense.
We remark that this is not always the case, especially for spectrographical
observations where the slit samples a number of internetwork points that is
usually much larger than that of network structures.
In Fig.  \ref{fig_cur}, as an example, we show  the light curves 
of different signatures for a NBP. 
Fluctuations are evident at each wavelength. The red and blue
wings of H$\alpha$ show a very similar temporal evolution, with simultaneous
intensity variations of the same magnitude for most NBPs. For other
signatures a direct comparison does not show any simple relation
between the variations at different atmospheric heights.

We must comment here on the fact that most authors, when studying the temporal
properties of network points, use a different method to determine the ``light
curves''. Basically, the network points are spatially identified using the 
temporal average of some suitable signature, and then the  temporal development
of  each {\it image pixel} belonging to the average
structures is considered and
analyzed. We believe that the method we adopted has several advantages: 
- 
Since the network points might move over the course of time, they do not
always correspond spatially to the structures identified on the average maps.
Our method guarantees that the structure is properly followed in time, avoiding the
loss of relevant pixels, or the inclusion of spurious ones; -
If the
magnetic structure giving rise to the network point inclines with height, a set
of pixels identifying the NBP at a given wavelength might not well
represent the same structure at a different wavelength. This is especially
relevant when performing comparisons between different atmospheric layers, for
example in the phase difference analysis of Sect. \ref{s_phase}.
We overcame this problem by selecting areas large enough to include the
NBPs at each wavelength and each time, as explained earlier.

Computing the light curves as an average over a given area 
implies the assumption of a spatial coherence over the whole area 
(of about 3\arcsec$\times$3\arcsec, equivalent to a spatial frequency of about 3
Mm$^{-1}$) competing to each
NBP. This seems a reasonable assumption because we do not see any significant
inhomogeneities within the single structures.

\section{Power spectra}
\label{s_power}

A search for possible periodicities in the fluctuations of the NBPs light 
curves was performed using temporal power spectra.
Before computing the power spectra, the light curves
were detrended using a smoothing window of 600 s.
 A check on this procedure
showed that changing the smoothing window between 360 s and 840
s affected the power at frequencies lower than 1.2 mHz, but without changing
the frequency of the peaks. The power at higher frequencies remained
unaffected. 

\begin{figure*} 
\resizebox{12cm}{!}{\includegraphics{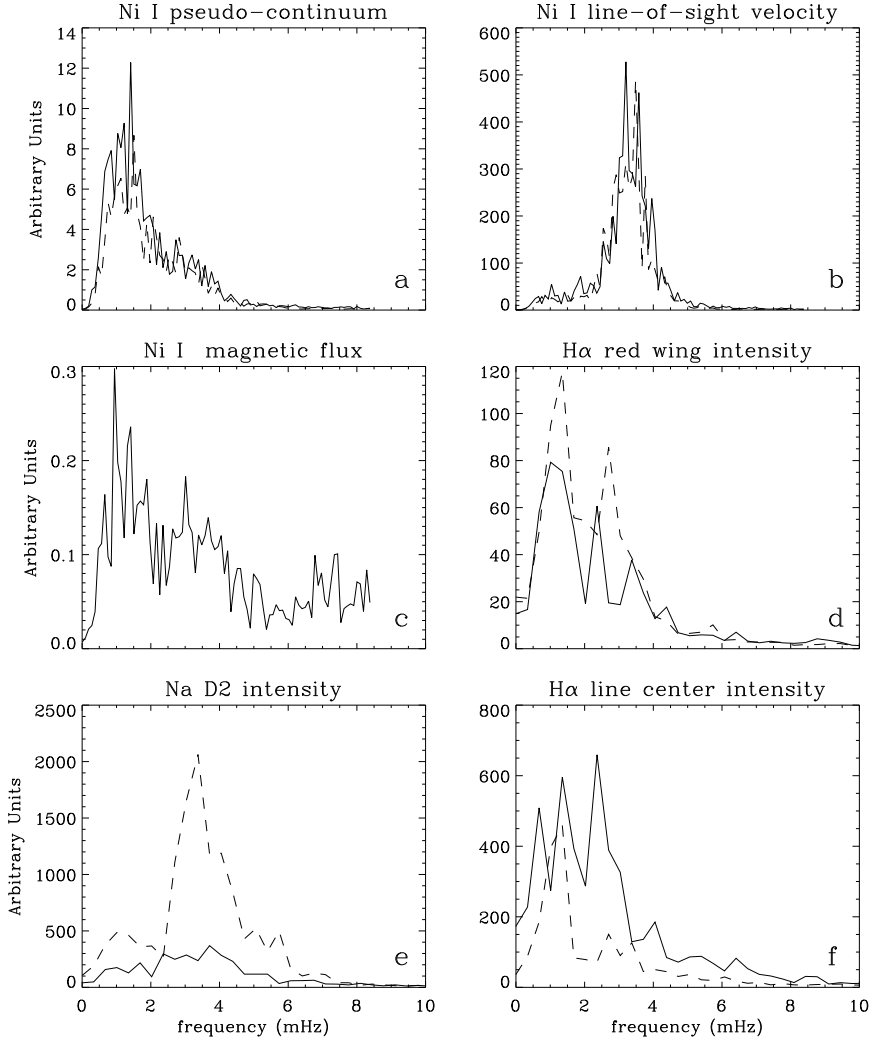}}
\caption{
Power spectra in arbitrary units, 
averaged for NBPs (solid) 
and internetwork regions (dashed). 
The 50\% confidence limit is given by the range 0.8$-1.5\times$ the actual power
values; for clarity error bars are not plotted in the figure.
} \label{figpower}
\end{figure*}

A power spectrum was computed for each light curve of the 11 NBPs
and of the 11 internetwork areas.
To analyze the differences between these two atmospheric components,
we averaged separately the power spectra over all of the NBPs and 
over all of the quiet regions.  
In Fig. \ref{figpower} we show some of these averaged curves.
It must be remembered that
the NSO and MDI observations have
different temporal coverage (50 and 180 minutes, respectively) and
temporal resolution (12 s and 60 s), so that lower temporal
frequencies are better represented in the NiI series, while frequency coverage
extends to higher frequencies for ground-based data. 
However, the power at $\nu \ge 8$ mHz in all the ground-based signatures is
due essentially to noise (Fig. \ref{figpower}), hence the analysis can 
be limited to the frequency coverage of the NiI
observations, 0 - 8 mHz.

We describe here the power spectra characteristics from lower photospheric 
signatures to higher chromospheric ones. 
Intensity fluctuations may be plausibly interpreted as
temperature fluctuations for photospheric LTE signatures 
such as Ni I or the \Ha~wings, formed over depths where the velocity
gradient is small. In these cases, the intensity fluctuations directly
reflect fluctuations of the source function (the Planck function) and
hence of temperature.
In the center of chromospheric lines as Na$D_2$ or \Ha \ the NLTE
effects are important and
the intensity fluctuations are the response to temperature, density and
even velocity changes (Cram 1978) and then are a
sort of average of the variation of the state variables of the chromosphere.

\subsection{Photospheric signatures}
\label{s_power_phot}

In Fig. \ref{figpower} a, b, d we show the power spectra of Ni I
pseudo-continuum intensity, Ni I velocity and \Ha \ red wing intensity.
The power spectra computed for white light intensity (not shown in figure) 
and for the NiI
pseudo-continuum are very similar, and we
are confident that we can compare the two series of observations, 
even if their frequency resolution is different.
The \Ha \ blue wing spectrum (not shown in figure) has a 
similar behaviour but a lower amplitude with respect to the red wing
(see  Table \ref{tabpow}).
A value smaller than 1 for the ratio of the power in the blue and red wings
of \Ha~ is consistent with the observations of Bertello (1987), that found
the same trend for the power of velocity 
oscillations in the wings of photospheric lines formed at 
heights lower than 150 km.

As is well known, the distribution of power for the photospheric
velocity fluctuations is rather different from the one of 
pseudo-continuum intensity
fluctuations (Fig. \ref{figpower} a-b).
The velocity power is concentrated around
the range of frequencies corresponding to the 5-minutes oscillations. 
The pseudo-continuum 
power spectrum peaks at low frequencies, around 1.5 mHz 
and then show a decay that might indicate the stochastic character of the
granulation intensity variation, as already reported for the first time 
in Noyes (1967).

As a global characteristic, power spectra computed in photospheric
signatures do not show any significant difference between network and
internetwork structures within the 50\% confidence limit (Fig.
\ref{figpower} a, b, d).  
This result
is consistent with previous spectral observations by several authors
(\cite{deu90}; \cite{kul92}; Lites et al. 1993) 
that analyzed both intensity and velocity oscillations in the
photosphere for network and internetwork features.

The power spectrum of the magnetic flux variations averaged over
the NBPs is shown in Fig. \ref{figpower} c. Internetwork areas are 
not considered because 
the noise in the magnetic flux measure is too high for
a reliable determination of fluctuations. Significant peaks are visible 
at low frequency (around 1.5 mHz), indicating long term evolution of the
magnetic field, and around 3.5 mHz corresponding to the 5 minutes oscillations.
A signal at the latter timescale might represent the magnetic response to
oscillations already present in the photosphere and be of
importance in the context of generation and dissipation of MHD waves in 
the solar atmosphere (\cite{ulr96}).
Observations of flux variations in small magnetic structures are
scarce in the literature, but we can compare this result with those
presented by Norton et al. (1999), that used a similar set of MDI data
obtained in the area of a big sunspot.
They found a significant peak  near 5 min only for 
structures whose magnetic flux density exceeded 600 G, while the points we
analyzed had a maximum value of about 300 G.

\subsection{Chromospheric signatures}
\label{s_power_chrom}

The intensity  power spectra computed in chromospheric 
signatures display
strong differences between NBPs and internetwork as shown in 
Fig. \ref{figpower} e - f. 
We will analyze in detail these differences, keeping in mind that the regime
of oscillations changes with height in the chromosphere.

First of all we notice that the power spectrum of internetwork intensity 
fluctuations in NaD$_2$ shares some
characteristics with the photospheric  NiI velocity power spectrum
rather than with the one of Ni I intensity. In particular the
strongest peak appears around 3.5 mHz, while the enhanced low frequency component, 
typical of photospheric intensity power spectra, is lacking.
This characteristic  could be explained if the intensity
fluctuations in the NaD$_2$ line center were related more to velocity than to
temperature perturbations. This might be indirectly confirmed by the results of
Pall\'e et al. (1999) in their study of the current
performances of the GOLF experiment on SOHO. In determining the 
relative contributions of velocity and intensity signals
to the intensity variations measured in the blue wings ($-$100
m\AA) of the sodium doublet, they conclude that the effect due to ``pure'' 
intensity changes is only 14\% that of velocity changes
for the {\it p} - mode frequency range. 
Since the width of the filter used for our observations includes that same
portion of the line wing, this conclusion might apply, at least partially,
also to our case.

Comparing network and internetwork power spectra for \Na, we see that
the power of the NBPs is smaller than the corresponding power for
internetwork at each temporal frequency.
In particular, even if both NBPs and internetwork points show a maximum
around 3.5 mHz, the power at this frequency is almost an order of magnitude
smaller in the NBPs.
(Fig. \ref{figpower} d suggests that this effect might be already 
present in the wings of H$\alpha$,
although its amplitude is not large enough to give an unambiguous
result). 
This suppression of power in the NBPs indicates that the
presence of the magnetic field in some way perturbs and
reduces the oscillations at low chromospheric levels, especially in the
$p$-mode range. A compression of power in magnetic structures at
frequencies below 7-8 mHz, for lines formed at similar heights, 
has not been reported by other authors. Only
Al et al. (1998) observe a similar effect, but much smaller,
for the power spectrum of velocity fluctuations measured in the center
of \Na~with a narrowband filter (30 m\AA).  We think that the stronger effect
seen in our observations is real because our analysis
selects the horizontal scale typical of chromospheric network to
determine the light curves and the power spectra (see Sect.
\ref{s_lcurves}), and is therefore more
suitable to outline characteristics and differences on the same
horizontal scale for NBPs and internetwork areas.

The power spectrum computed for the intensity of \Ha \ center is shown in
Fig. \ref{figpower}  f. For both NBPs and internetwork
the power distribution peaks at low frequencies ($\nu \leq 3$ mHz),
without any relevant peak at the $p$-mode frequencies.
No enhanced power for ``3-minutes'' oscillations  ($\nu
> 5$ mHz), is detectable in the internetwork. 
This is consistent with observations in the \Ha \ center by Cram (1978)
and in the Ca II - H3 by \cite {lit93} showing a power peak in the ``3
minutes'' range only for velocity power spectrum.

The spectrum of NBPs in \Ha~line has a power higher than that of the 
internetwork
areas at each frequency, reversing the effect present in the
\Na~line, and shows three well separated  peaks reminiscent of the peaks 
observed by Lites et al. (1993).
The more relevant peak is around 2.2 mHz (``7-minutes'' oscillations)
and is lowered about a factor 6 in the internetwork. 
We cannot judge on the relevance of the peak around 1.3 mHz, since its amplitude
is heavily affected by the smoothing window, as described in 
Sect.\ref{s_power},
and we cannot consider real the peak at 0.6 mHz, because it is related
to the time interval of our observations.
However, the increasing power at low frequencies in the spectrum of
NBPs suggests that
the r\^{o}le of the magnetic field in the oscillations, detectable 
at high chromospheric levels, is certainly different from the one at
lower levels and strenghtens  the hypothesis of magnetic-hydrodynamic 
waves present at these high levels.

Characteristics of power spectra for network and internetwork are
summarized in Table \ref{tabpow} for the two relevant frequency windows 
1.5 - 2.5 mHz and 3 - 4 mHz. 
\begin{table}
\caption{\label{tabpow}Power values in arbitrary units for various observed 
features. A range of values is reported when variations are large.}
\begin{flushleft}\scriptsize
\begin{tabular*}{7.8cm} %
{@{\extracolsep{\fill}}lllll}
frequency range& \multicolumn{2}{c}{  1.5 - 2.5 mHz }  & 
\multicolumn{2}{c}{  3.0 - 4.0 mHz }  \\ 
 & & & &\\
Observed features & NBP & Internet. & NBP & Internet.\\ 
 & & & &\\
Pseudo-continuum (Ni I)   & 8 - 2 &  8 - 2 & 2 & 2\\
Velocity (Ni I) & -- & -- & 400 & 400 \\
Magnetic Flux & 0.15 - 0.1 & -- & 0.13 & -- \\
H$\alpha$ red wing & 80 - 40 & 100 - 60 & 30 & 30 \\
H$\alpha$ blue wing & 50 - 30 & 75 - 45 & 20 & 25 \\      
NaD$_2$ & -- & -- & 300 & 2000 \\
H$\alpha$ center & 650 & 100 & -- & -- \\

\end{tabular*}
\end{flushleft}
\end{table} %

\section{Phase difference and coherence spectra}
\label{s_phase}

In order to look for propagation characteristics of 
waves at different heights in the atmosphere, 
we computed phase difference ($\Delta\Phi$) and phase coherence ($C$)
spectra for many signature pairs  (\cite{tho95}).
We exclude in this analysis the \Na~signature because, if the measured
intensity oscillations are due essentially to velocity oscillations, 
we cannot establish the direction of the motion and hence
assign the correct value to the phase difference.

Following Edmonds \& Webb (1972), 
we adopted a smoothing width of about 1.5 mHz in
the Fourier domain for the computation of phase and coherence spectra. 
Frequencies smaller than 0.7 mHz hence do not convey any
significant information. Taking into account 
our smoothing width, a coherence smaller than $\sim$0.5 implies that 
phase differences at those frequencies are completely unreliable.

NSO and MDI data were analyzed independently, using their own frequency
resolution and coverage. 
Finally, the study was performed separately for internetwork
and network areas, to distinguish between features with different magnetic
characteristics. We describe the characteristics of
the spectra from lower photospheric layers through higher chromospheric ones.

\begin{figure*}
\vspace{1.5cm}
\resizebox{12cm}{!}{\includegraphics{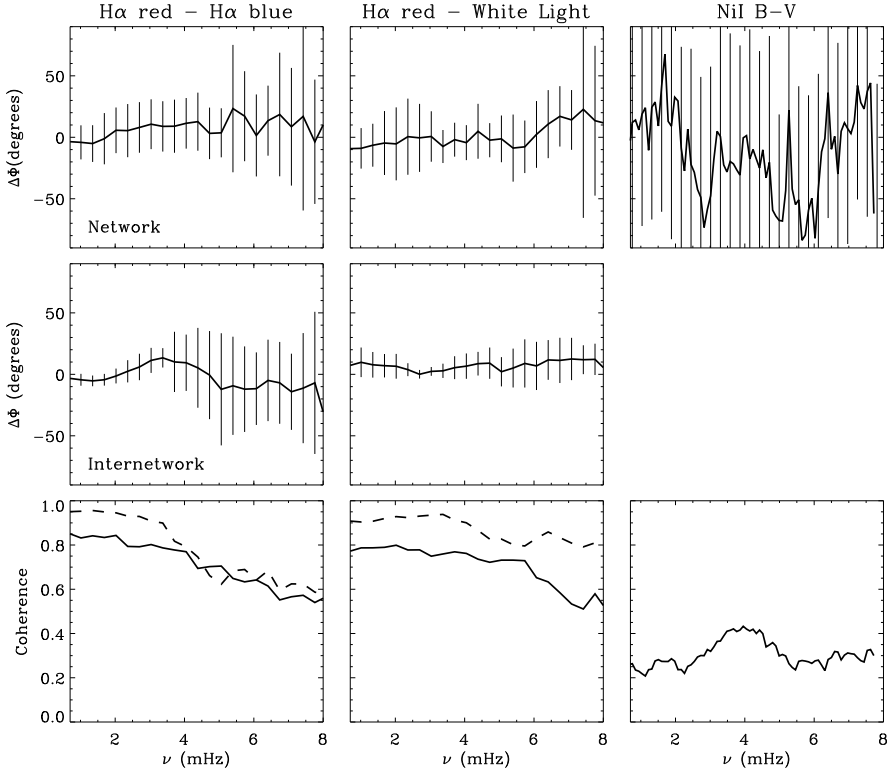}}
\caption{\label{fig_pho} Top and mid  panels show the phase difference I $-$ I, 
as a function of frequency, of photospheric signatures averaged for network 
and internetwork regions. 
Vertical bars, representing 1$\sigma$ error associated with the mean phase value,
are clipped if higher than the y-axis limits.
The bottom panels show the phase coherence spectra (computed with a smoothing 
width of 1.5 mHz, see text) averaged for network (solid line) and 
internetwork (dashed line) regions.} 
\end{figure*}

In Fig. \ref{fig_pho} left and central column,  we show intensity (I$-$I) phase difference and 
coherence spectra for the pairs \Ha~$+$ 1.5 \AA~/\Ha~$-$ 1.5 \AA, and
\Ha~$+$ 1.5 \AA~/ white light, originating at photospheric levels. 
We use here the white light signal (and not the NiI pseudo-continuum) in order to
keep the maximum possible frequency resolution. 
Phase difference and coherence spectra between magnetic 
flux density and velocity (B$-$V) for the NBPs are shown in Fig.
\ref{fig_pho} right column. 

\begin{figure} 
\resizebox{8.8cm}{!}{\includegraphics{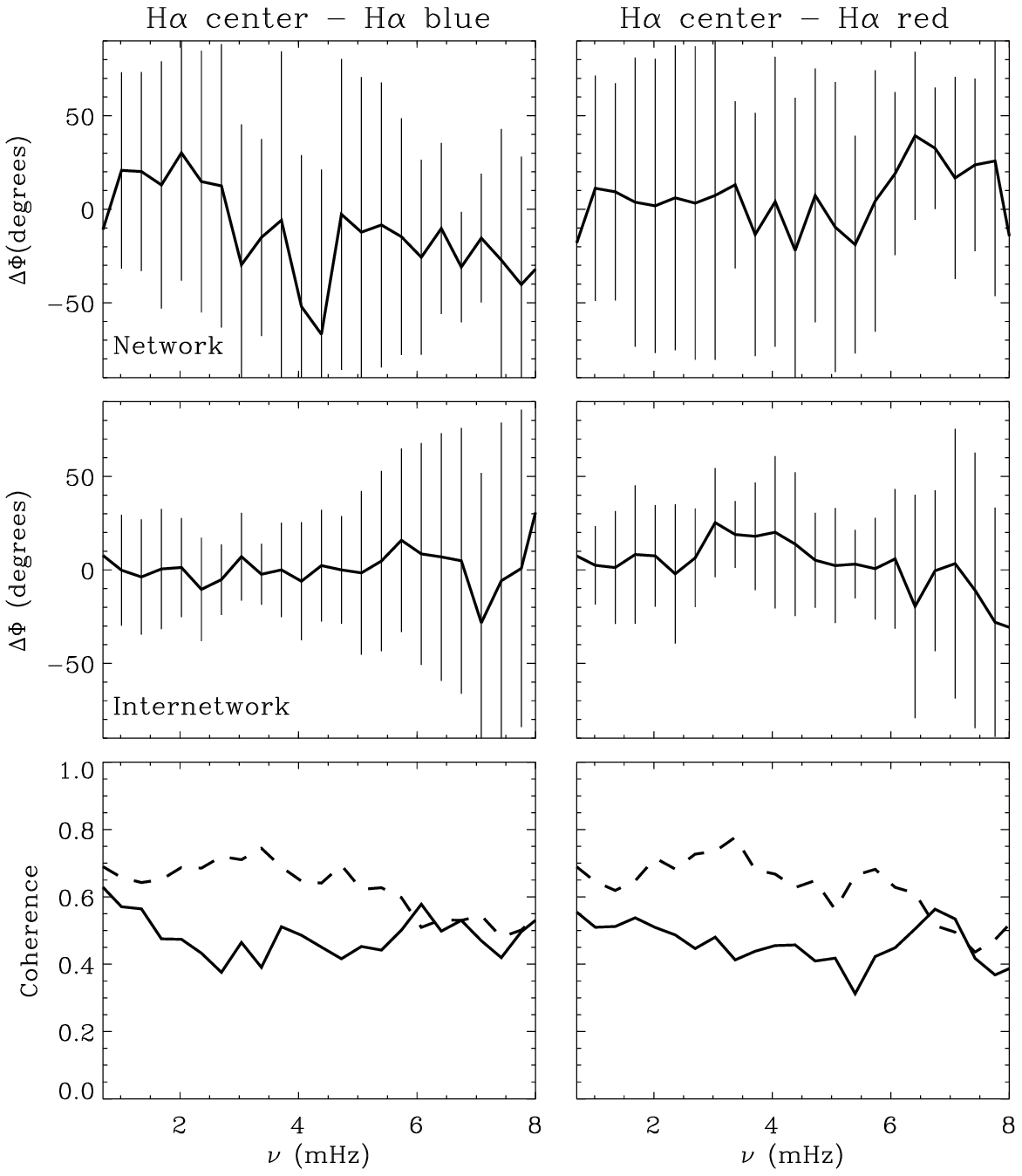}}
\caption{\label{fig_chr} Top and mid panels show the phase difference I $-$ I,
as a function of frequency,
for the pairs \Ha~center / \Ha~$-1.5 $ \AA \ and \Ha~center /
\Ha~$+1.5 $ \AA \ averaged for network and internetwork regions.
Vertical bars, representing 1$\sigma$ error associated with the mean phase
value, are clipped when higher than the y-axis limits.
The bottom panels show phase coherence spectra (computed with a
smoothing width of 1.5 mHz, see text) averaged for network (solid line) and 
internetwork (dashed line) regions.}
\end{figure}

To search for the relationship between the oscillations present at
chromospheric and  photospheric levels we
computed the I$-$I phase difference and coherence
spectra separately for the pairs \Ha~center / \Ha~$-1.5 $ \AA \ and \Ha~center /
\Ha~$+1.5 $ \AA, shown in Fig. \ref{fig_chr}. 
 At each considered atmospheric level, a general characteristic is that 
the coherence for NBPs is
smaller than for internetwork, hence the phase values for NBPs are
more uncertain.

We examine different signature  pairs separately in the two frequeny
intervals 1.5 - 2.5 mHz and 3 - 4 mHz, disregarding the features with
negligible power, and we
summarize in Table \ref{tab_pha} the phase and coherence values for each 
pair. 

\begin{table}
\caption{\label{tab_pha} Phase difference $\Delta\Phi$ in degrees and coherence {\bf C } 
for the considered pairs }
\begin{flushleft}\scriptsize
\begin{tabular*}{7.8cm} %
{@{\extracolsep{\fill}}llllll}
frequency range& &\multicolumn{2}{c}{  1.5 - 2.5 mHz }  & 
\multicolumn{2}{c}{  3.0 - 4.0 mHz }  \\ 
 & & & &\\
 Observed pairs & & NBP & Internet. & NBP & Internet.\\ 
& & & &\\
\Ha~$+$ 1.5 \AA~/\Ha~$-$ 1.5 \AA & $\Delta\Phi$ & 0 & 0 & 7 & 10\\
\Ha~$+$ 1.5 \AA~/\Ha~$-$ 1.5 \AA & {\bf C } & 0.85 & 0.95 & 0.8 & 0.9\\
& & & & &\\
\Ha~$+$ 1.5 \AA~/\ WL & $\Delta\Phi$ &  $-8$ - 0 & 10 - 5 & 0 & 0 \\
\Ha~$+$ 1.5 \AA~/\ WL & {\bf C } & 0.8 & 0.95 & 0.75 & 0.95 \\
& & & & &\\
Magnetic flux / Velocity & $\Delta\Phi$ & -- & -- & $-20$ & -- \\
Magnetic flux / Velocity & {\bf C } & -- & -- & 0.4 & -- \\
& & & & &\\
H$\alpha$ center /\Ha~$-$ 1.5 \AA & $\Delta\Phi$ & 0 & 0 & -- & --   \\
H$\alpha$ center /\Ha~$-$ 1.5 \AA & {\bf C }& 0.45 & 0.72  & -- & --  \\
& & & & &\\
H$\alpha$ center /\Ha~$+$ 1.5 \AA & $\Delta\Phi$ & 0 & 0 & -- & --  \\
H$\alpha$ center /\Ha~$+$ 1.5 \AA & {\bf C }& 0.5 & 0.65 & -- & --  \\

\end{tabular*}
\end{flushleft}
\end{table} %

\subsection{Low frequency (1.5 - 2.5 mHz)}
For both internetwork and network points the I$-$I phase difference 
between the \Ha~red and blue wings is  
0\degr.
The  two signals should be in  phase if the observed oscillations are 
due only to temperature and in antiphase if due to velocity.
We can then state that the observed oscillations at low frequencies
are essentially due to temperature oscillations.

It follows from the previous considerations that
the analysis of the phase difference spectra between the \Ha~red wing 
and white light is essentially a study of the correlation  between
temperature oscillations at slightly different levels 
 ($\Delta h \leq 100$ km in the quiet average photosphere). 
For internetwork areas the extremely high coherence makes the 
$\Delta\Phi$ value (5 - 10\degr)  highly significant (see Table
\ref{tab_pha})
 and strongly suggests that the observed power is due to oscillations. 
A positive value of $\Delta\Phi$
between two layers with decreasing heights, indicates the
presence of waves  directed radially inward. For the internetwork areas,
free from magnetic fields, at the spatial frequency of
about 3Mm$^{-1}$ used in our analysis and within the
considered frequency range, these waves might be interpreted as gravity waves
(\cite{tho95}). This same interpretation has been adopted for
internetwork by Rutten (2000), who reported
a similar value of $\Delta\Phi$  
 between the intensity of two continuum levels observed by TRACE. 
Downward directed gravity waves had been proposed earlier
by Staiger et al. (1984) to explain similar 
phase differences for velocity signatures at photospheric levels.

At chromospheric levels, the most significant peak in the power 
spectrum of the \Ha \ center intensity appears at 2.2 mHz for NBPs
and is not related to the oscillations present
at photospheric levels ( {\bf C} $= 0.45$ between \Ha~center /\Ha~$+1.5
$ \AA, see Table \ref{tab_pha}). This means that 
at chromospheric levels the NBPs experience
a new regime of oscillations that seem to be independent from what
happens in photosphere.

\subsection{$p$-mode frequency (3.0 - 4.0 mHz)}
 In the $p$-mode range of frequencies, for 
internetwork there is a 10\degr phase lag between the two \Ha~wings
(see Table \ref{tab_pha}) that can be due
to different coupling of velocity and intensity fluctuations in the two
wings. This effect has been extensively studied for 
photospheric lines by Cavallini et al. (1987) and  Alamanni et al.
(1990).

As described in Sect. \ref{s_power_phot}, the presence of
a peak at these frequencies %
in both the magnetic flux 
and velocity power spectra could betray the presence of MHD waves in the
network structures.
In the limit of ideal MHD, a definite phase relation between velocity and 
magnetic field variations is expected as signature of Alfv\`en waves 
($\Delta\Phi =
0\degr$) or magnetoacoustic waves ($\Delta\Phi = 90\degr$, with v leading
B. See \cite{ulr96}).
However, we find a very low coherence ($<$ 0.4) at all
frequencies in the NBPs, i.e. the magnetic fluctuations are not related to the 
velocity ones, at least with the present
sensitivity and resolution.

\section{Discussion and conclusions}
\label{s_concl}

The observations presented in this paper allowed us to define
the characteristics of network bright points 
at different atmospheric heights, and to compare them with those of the
surrounding internetwork areas.  We improved on the existing statistics using a
good-sized sample of NBPs, and the same number of ``test'' internetwork
areas, defined in a comparable way. The method we adopted to study
the temporal evolution of NBPs insures that each bright structure is
properly followed in time and position at each height.
In fact, the evaluation of the light curves and their properties
after a spatial averaging over a
well defined area guarantees that we are studying the same NBP at all
heights, and
avoids the problem (first pointed out by \cite{lit94}) of a possible
structure displacement  due to the magnetic field inclination.
Given the characteristic horizontal size of the NBPs,
the analysis and the
comparison of power spectra and phase differences concern the 
propagation of waves pertaining to a horizontal wavenumber of about 3
Mm$^{-1}$.

The quasi-simultaneous series of \Na~images and of 
MDI maps allowed us to establish for the first time a correspondence between
\Na~bright network and magnetic network at high spatial and temporal
resolution. A correspondence between bright chromospheric structures (Ca
II, Ly$\alpha$, Mg I and UV continuum) and magnetic structures had been
observed before, but not at this high temporal resolution. We also
established for the NBPs a quantitative relationship between the Na 
excess and the corresponding absolute value of magnetic flux density. This
relationship is best expressed by a power law with an exponent very close
to the one found by 
Schrijver et al. (1989, 1996) for the Ca II - K excess, and indicates that the
emission in \Na~may be used as a proxy for the magnetic flux
density.

The NBPs considered in this work have the
following properties: - are bright in the Ca II wings and in the Ca II 
K$_2$ peaks; - are visible in the \Na~ images for about 1 hr; - coincide
spatially with the magnetic structures; - are nearby or within a lower
activity region. 
The general characteristics found for these NBPs do not
differ from the ones derived in absolutely quiet regions
(\cite{deu90}, \cite{lit93}).

 Our results referring to photospheric and chromospheric properties are so
summarized: 

{\it At photospheric levels:} No difference is detected
between network and internetwork power spectra, either in intensity or in
velocity,   within the limits of sensitivity and accuracy of the
instruments used for this work. The phase difference spectra between
photospheric signatures  in general do not show different characteristics
for network or internetwork. However, when analyzing the phase difference
between \Ha \ red wing and white light images($\Delta h \leq 100$ km),  we
find $ 5\degr \leq \Delta\Phi$ $\leq 10\degr$ \  in the frequency window 
1.5 -  2.5  mHz and in the internetwork.
(The $\Delta\Phi$ value is more uncertain in the network, due to a lower
coherence value). A phase lag of this amplitude and sign
is usually considered a signature of
gravity waves directed radially inward. A possible explanation for 
their origin might be
sought in recent  models of convection, described as a non-local process  
driven by
{\it cooling} at the solar surface rather than by heating from the lower
layers (Spruit, 1997). One can imagine that the downward flowing cooled
plasma can trigger some inward directed waves, and hence justify the fact
that the external layers ``lead'' the deeper layers. The general inhibition
of convection in magnetic structures might be the reason for the lack of
this signature in network points.

The power spectrum of the magnetic flux variations in NBPs shows a small but
significant peak around 3 mHz, that could be related to a
``transformation'' of acoustic waves into MHD waves. However, 
the phase difference and coherence
spectra between magnetic  flux and velocity (B$-$V) for the NBPs 
indicate a very low correlation between the two signals so we cannot
conclude anything on the presence of MHD waves within the network
points.

{\it At chromospheric levels:} Network and internetwork areas
have a rather different behaviour in the power spectra. We do not see
any evidence for the typical chromospheric period of 3 minutes (but it
must be reminded that they are best seen in velocity variations rather than
intensity).
In the low
chromospheric levels, where \Na~originates, the NBPs power
spectrum is compressed at all frequencies if compared to the 
internetwork, while in the high chromosphere,
where \Ha \ originates, the power of NBPs is higher than the one of
internetwork. 
This opposite effect may be an indication that the magnetic field 
disturbs and reduces the
amplitude of oscillations already present in the low chromosphere while
it assumes a leading r\^{o}le in the high chromosphere.

In the layers contributing to the \Na~emission it
seems that the oscillations present in network points change regime with
respect to both the photosphere and the high chromosphere and we
think that it would be important to perform observations of
NBPs in tha Na line, with high spectral resolution.
Unfortunately we cannot analyze the phase
difference spectrum for \Na~intensity fluctuations with respect to  others 
formed at different layers, since the \Na~intensity fluctuations, measured 
with the UBF filter (FWHM$=0.2$\AA), are more related to velocity than to 
temperature fluctuations (see Sect \ref{s_power_chrom}).

The power spectrum of \Ha \ intensity in NBPs has the more relevant
peak at 2.2 mHz, but this signal is not correlated with the photospheric
fluctuations, as indicated by the very low coherence measured at all
frequencies between the \Ha \ core and the blue and red wings.
We can then confirm, using a larger sample of NBPs, the presence of 
the peak found by Lites et al. (1993) around 2 mHz in the power spectrum of
K3 velocity fluctuations for one network point. Kalkofen (1997) and
Hasan \& Kalkofen (1999) proposed  an explanation for this peak in
terms of transverse  magneto-acoustic waves in magnetic flux tubes,
excited by granular
buffeting in the solar photosphere. In their model the low coherence between
photospheric and chromospheric signatures could be explained by a 
partial conversion of the transverse waves to longitudinal modes in the
higher chromosphere.

A general result of our analysis, valid from the low photosphere to the
high chromosphere, is
that the NBPs always show  a coherence lower than the internetwork,
pointing out that the presence of the magnetic field changes
the propagation regime of waves  with respect to the non-magnetic
regions.

\begin{acknowledgements}
The authors are indebted to the NSO/SP-DST staff for the generous
telescope time allocation and the unvaluable help during the
observations.
The authors express their thanks to the 
MDI team (P.I. P.H. Scherrer) for the efficient support
during the observing run and their dedication in operating this
instrument. MDI is part of SOHO, the Solar and Heliospheric Observatory,
mission of international cooperation between ESA and NASA.
We woud like to thank Thomas Straus for his comments during the
preparation of this paper and for fruitful discussion on the phase
difference problem. 
\end{acknowledgements}

\appendix
\section{A relationship between Na excess and magnetic field}
\label{s_app}

A quantitative relationship between the CaII K line core intensity and the 
absolute value of the magnetic flux density has been clearly established by
several studies (\cite{sku75};  \cite{schri89}; \cite{nin98}). 
The best agreement with the  data is 
given by a power law relation with an exponent of 0.6 (\cite{schri89}).  
Given the excellent spatial coincidence of the sodium network points and the
magnetic structures present in our data (see Sect. \ref{s_magd2}, 
we searched for such a quantitative relationship also for this case.

\begin{figure}[h]
\resizebox{8.8cm}{!}{\includegraphics{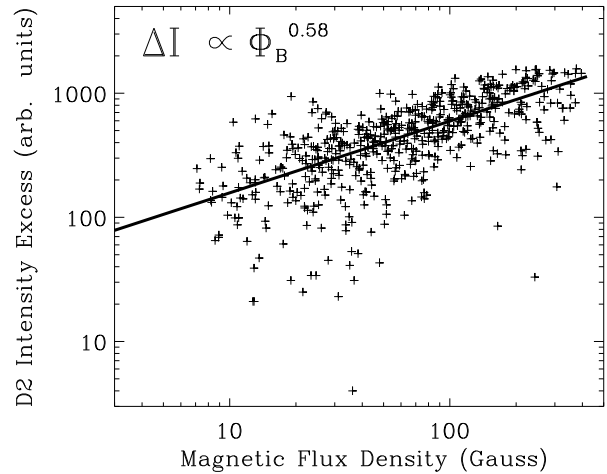}}
\caption{\label{scatter}Scatter diagram of the NaD$_2$ intensity, after
subtraction of a threshold of about 10$^4$ (see text), 
vs. the absolute longitudinal
 magnetic flux density. All the 11 analyzed NBPs are plotted.
Solid line is  the power law fitting described in the text.}
 \end{figure}

Fig. \ref{scatter} shows a scatterplot of the sodium ``excess'' for
the  11 NBPs, vs. the corresponding absolute values of magnetic flux density.
As done by Schrijver et al. (1989, 1996) for the CaII K emission, 
we subtracted to the NaD$_2$ intensities a threshold equal to
the average intensity in the quiet areas ($10^4$ in arbitrary units).
The graph was obtained plotting the temporal average of both  the
sodium intensity and magnetic flux. It does not display, hence, any temporal
variation of either quantity, but only a general trend among the persistent
structures. Five-minutes oscillations do not play a role in this  relationship,
nor do they contribute to the scatter of the points. 
Saturation, as reported by Schrijver et al. (1989) is not apparent in our data,
but it must be remarked that the maximum value of the magnetic flux density
for the average quantities is below 400 G,  the ``critical'' value
indicated by those authors.

The sodium excess is best
fitted by a power law of the type $\Delta$I $\propto  \Phi^\beta$, with
$\beta$=0.58 $\pm$0.1. The exponent is very close to 
$\beta$=0.6$\pm$0.1 found by Schrijver et al. (1989, 1996) for the CaII K
excess. These results indicate that the emission 
in the center of the NaD$_2$ line 
is also a good proxy for the magnetic flux density and, at least for values of
magnetic flux density smaller than a few hundred G, its use is 
equivalent to that of the CaII K emission. 

\end{document}